\documentstyle[12pt]{article}
\date{}
\newcommand{\beq}{\begin{equation}}
\newcommand{\eeq}{\end{equation}}
\newcommand{\beqn}{\begin{eqnarray}}
\newcommand{\eeqn}{\end{eqnarray}}
\newcommand{\dst}{\displaystyle}
\newcommand{\fr}[2]{\frac{{\dst #1}}{{\dst #2}}}
\newcommand{\eq}[1]{eq. (\ref{#1})}

\title {INITIAL PARTICLE INSTABILITY IN MUON COLLISIONS}

\author{I.F.~Ginzburg\\
Institute of Mathematics, Novosibirsk, 630090, Russia\thanks{
Work is supported by grants of Russian Foundation of Fundamental
Investigations and European association INTAS (INTAS -- 93 --
1180).}\\ e-mail: ginzburg@math.nsk.su}

\begin{document}

\maketitle

\begin{abstract}
I consider the process $\mu^+\mu^-\to e\bar{\nu}W^+$ in the case
when the effective mass of the $(e\bar\nu)$ system than the muon
mass. In this case the momentum transferred from the initial
muon to the $e\bar\nu$ system (the virtual neutrino momentum) can
be both time--like and space--like. Since the path of integration
over $k^2$ goes through a pole at $k^2=0$, it gives a divergent
cross section.

In the ideal case of large enough beams this divergence
disappears if the finite width $\Gamma$ of the initial muon is
taken into account. The obtained cross section corresponds to the
flux of equivalent neutrino, which coincides with that of muon
(with some energy distribution). 

In practice, the effect of final size of the muon beam reduces
this cross section very strong, and the effect is hardly
observable.
\end{abstract}

Recently, muon collisions have been proposed as the next step
for high--energy colliders (see e.g. \cite{Palm}). This idea
provides a problem:

{\em To find (if possible) the point where the muon collisions
differs substantially from the electron ones.} My first
impression was: I found this point; it relates to the muon
instability, which gives really new option to consider muon
collider as neutrino collider simultaneously. The subsequent
studies shows that the discussed effect is small in practice, the
basic problem has a negative solution for the muon collider
project. Nevertheless, the discussed problem seems to be
important for the particle theory. The first part of the
discussion below reproduces the paper \cite{Gin}, the final
result for the "realistic" beams is given from ref. \cite{MSer}.

\section{The problem}

We discuss the effects of the muon instability for the process
\beq
\mu^-(p_1)\mu^+(p_2)\to e(q_1)\bar{\nu}(q_2) W^+(p_3).\label{process} 
\eeq
where the momenta of the particles are shown in brackets. We use
the following notation:\ $M$ is the $W$ boson mass, $m$ is the
muon mass, $s\equiv 4E^2 =(p_1+p_2)^2$; $\;x=M^2/s$;
$\;q=q_1+q_2, \quad k=p_1-q\equiv p_3-p_2$ and we neglect the
electron mass.

1. We study the specific kinematical region where the effective
mass of the $e\bar{\nu}$ system is less than the mass of the
muon, $q^2<m^2$. In this case the transferred momentum $k$ can be
time--like, the maximal value of $k^2$ is positive:
\beq
k^2\leq t_{max} =\fr{x}{1-x}\left[m^2 (1-x)-q^2\right] >0.
\label{bound}
\eeq
With the increase of the total transverse momentum of the
produced system, this momentum--transfer becomes space--like,
$k^2<0$. Therefore the integration over this transverse momentum
(at fixed $q$) goes through the point $k^2=0$. 
 
The main contribution to the cross section in this region is due
to the diagram of Fig. 1, where the neutrino t--channel exchange.
It gives a factor $(k^2)^{-2}$ in the matrix element squared. The
standard integration over $k^2$ results in a divergent cross
section in this case!  ( This longstanding problem has got
no satisfactory solution till now (see \cite {Pai,Eden,Web}, for
recent review see \cite{Pfi}).)

\begin{figure}[hbt]
\begin{center}
\unitlength 1mm
\begin{picture}(80,50)
\put(0,40){\vector(1,0){15}}
\put(15,40){\line(1,0){10}}
\put(15,44){\makebox(0,0){$ \mu^-$}}
\put(15,9){\makebox(0,0){$ \mu^+$}}
\put(0,5){\vector(1,0){15}}
\put(15,5){\line(1,0){10}}

\multiput(25,40)(5,0){5}{\line(1,0){3}}
\put(40,40){\vector(1,0){3}}
\put(42,44){\makebox(0,0){$ W^-\;(q)$}}
\multiput(25,5)(5,0){7}{\line(1,0){3}}
\put(40,5){\vector(1,0){3}}
\put(45,10){\makebox(0,0){$ W^+$}}

\put(25,40){\vector(0,-1){20}}
\put(25,20){\line(0,-1){15}}
\put(32,23){\makebox(0,0){$ \nu\;(k)$}}

\put(50,40){\vector(3,1){12}}
\put(62,44){\line(3,1){8}}
\put(62,47){\makebox(0,0){$\bar{\nu}$}}
\put(50,40){\vector(3,-1){12}}
\put(62,36){\line(3,-1){8}}
\put(62,32){\makebox(0,0){$e $}}

\end{picture}
\caption{\em The diagram considered ($q^2<\mu^2$)}
\end{center}
\label{fig1}
\end{figure}
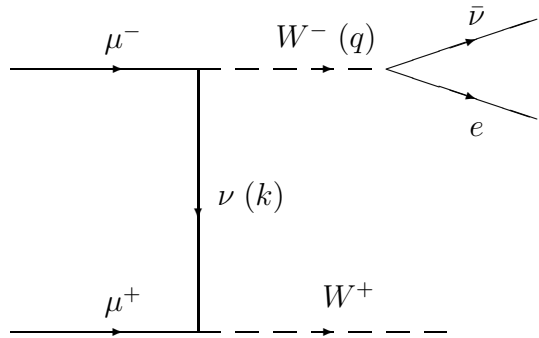

This paradox originates from the instability of the muon,
decaying into the $e\bar{\nu}\nu$ system: the point $k^2=0,
\;q^2 <m^2 $ is within the physical region for this decay. 

\section{Solution for the ideal case, large enough beams}

First, we neglect the beam size effects. In this case, the above
divergence is eliminated if one takes into account the fact that,
because the muon is instable, the wave function differs from the
standard plane wave. To obtain Lorentz covariant solution, we
start from the muon rest frame\footnote{ This way of deriving
$k^2_{new}$ was proposed by V.G.~Serbo.}. Here
\beq
e^{-imt/\hbar}\Rightarrow e^{-i(m-i\Gamma/2)t/\hbar}. \label{wf}
\eeq
In this frame the 4-momentum of $\mu^-$ is $\tilde{p}_1t =
(m-i\Gamma/2, 0,0,0)$. To obtain the energy of the produced
$e\bar{\nu}$ system, $\tilde{q}^0$, in this frame, we use the
simple kinematical relations $2p_1q=m^2+q^2-k^2,\;\;$
$2p_1q\equiv 2m\tilde{q}^0$, which give $\tilde{q}^0=
(m^2+q^2-k^2)/2m$.

The new value of $k^2$ is obtained from the relation $k^2_{new} 
\equiv (p_1-q)^2=m^2-im\Gamma +q^2 +i\Gamma\tilde{q}^0$. Using 
the value of $\tilde{q}^0$ given previously, we obtain \footnote{
In our approach, both the energy and
3--momentum of muon have imaginary parts in the lab. frame. The
idea of ref. \cite{Pai} looks like similar to that, discussed
here. The ansatz proposed can be written as $\gamma = m\Gamma
(1-x)$. To obtain this value, one should assume 3--momentum of
muon to be real in the lab. system, and calculate its energy with
complex mass. This approach is evidently not Lorentz covariant.}:
\beq
k^2\Rightarrow k^2-i\gamma \ ; \quad \gamma=\fr{m\Gamma (m^2-q^2)}
{2m^2}.\label{change}
\eeq

One can now calculate the cross section of the process in the
standard way. The calculation is simplified since one can
neglect the small quantities $\sim m^2/M^2$, $\Gamma/m$ in the
result. We finally obtain: 
\beqn
d\sigma&=&\fr{|{\cal M}|^2 dk^2 dq^2}{4(4\pi)^3 s(s-4m^2)} 
\fr{d\Omega^*_q}{4\pi}\fr{d\varphi}{2\pi};\nonumber\\
|{\cal M}|^2&=&\fr{(4\pi\alpha)^3}{M^2(\sin^2\Theta_W)^3} 
\fr{(2q_1 p_1)(2kq_2)}{k^4+\gamma^2}. \label{dsigma}
\eeqn
where $d\Omega^*_q$ is the solid angle element in the center of
mass of the produced $e\bar{\nu}$ system.

The subsequent angular integration is trivial. The integration
over $k^2$ gives $\pi/\gamma$ for any $q^2$, since the bounds of
the integration region are much higher than $\gamma$. The
integration over $q^2$ covers the region $q^2< m^2(1-x) $
(\ref{bound}).  This procedure is very close to the calculation
of the muon decay width.  We insert this width instead of the
corresponding combination of factors in \eq{dsigma}, and it
compensates the factor $\Gamma$ in the denominator; the final
result is then:
\beqn
\sigma&=&\fr{\pi^2\alpha}{s\sin^2\Theta_W } f(x)
\equiv 20xf(x)\mbox{ nb};\label{crsec}\\
 f(x)&=&4x(1-x)(2-x).\label{flow} 
\eeqn

This equation solves the discussed problem in the ideal case:
with the above prescription we obtain a finite cross section. It
is the final result for the description of hadron collisions with
fast decay of one of collided particles.

But the obtained quantities correspond to the integration over
the whole space--time irrespective to the size of interaction
region. The effective spatial scale of the considered phenomena
is $c\tau$ where $\tau =\hbar/\Gamma$ is the muon time of life.
In reality, the size of beam is much less, and only small
fraction of $\nu$ interacts. This very scale regularize the cross
section, the effect of muon instability manifests itself only in
the existence of region under interest with the possible
time--like momentum of exchanged neutrino, the imaginary part of
the muon mass become irrelevant to the observed phenomenon. 
 
\section{The finite size effect. Basic equations}

In the subsequent calculations we neglect the muon instability in
its mass, since considered sizes of bunch are very small. This
approach is justified by the finiteness of the observed result.

The calculations are based on the method, developed in refs.
\cite{GKPS} (see also \cite{Ter}). We should take into account, 
that the wave functions of muons in the initial beams are no
plane waves but wave packets with some distribution over momenta
(the effect of complex muon mass is hidden within this distribution):
\beq
|p_i>\to\int\fr{d^3 P_i}{(2\pi)^{3/2}}|P_i> \quad
(i=1,2).\label{wp} 
\eeq

When calculate cross section, we summarize over final states.
Therefore, we can use here arbitrary whole set of states, in
particular, plane waves $|q_i>,\;|p_3>$.

Therefore, the matrix element squared $|{\cal M}|^2$ is expressed
via the standard matrix elements in the momentum representation
as 
\beqn
|{\cal M}|^2&=&\int\fr{d^3 P_1 d^3 P'_1 d^3 P_2 d^3 P'_2}{(2\pi)^6}
M(P_1,P_2;q_1,q_2,p_3)M^*(P'_1,P'_2;q_1,q_2,p_3)\nonumber\\
&&\cdot\delta(P_1+P_2 - q_1-q_2-p_3)\delta(P_1+P_2-P'_1-P'_2).\label{m2}
\eeqn
We can write the identity:
$$
2\pi\delta(\sum P_i -\sum P'_i)=\delta(\sum {\vec P_i}
-\vec{\sum P'_i}) \int dt\exp[it(\sum \epsilon_i-\sum
\epsilon'_i)]\quad (\epsilon_i\equiv P_i^0).
$$ 
Then the phase averaging results in density matrices for the muons
in the beams:
\beq
<\Phi(P_i)\Phi(P'_i) \exp[it(\epsilon_i-\epsilon'_i)>= 
\rho(\vec{P_i},\vec{P'_i},t).\label{dens}
\eeq

Next, it is useful to go to the mixed representation of the
density matrix --- Wigner function $n(p,r,t)$:
\beq
\rho(\vec{P_i},\vec{P'_i},t)d^3 P_id^3 P'_i =\int n(\vec{p_i},
\vec{r_i},t)e^{2i\vec{l_i}\vec{r_i}}\fr{d^3 p_i d^3 l_i d^3 r_i}
{(2\pi)^{3/2}}\quad
\left(p_i=\fr{P_i+P'_i}{2},\,l_i=\fr{P_i-P'_i}{2}
\right).\label{wig} 
\eeq
In the quasi--classical limit, which realized for particles in beam,
this Wigner function coincides with the density in the phase
space. That is the point, in which known distributions of
particles within beams enter into the result. 

We see, that the identical final state is obtained from the
different (in plane wave language) initial states of all initial
particles and they give different values of transferred momenta
for the diagram and the conjugated one. Therefore, when calculate
probability ${\cal P}\propto|{\cal M}|^2$, we obtain instead of
(\ref{dsigma})
\beq
{\cal P} \propto \int n(\vec{p_1},\vec{r_1},t)n(\vec{p_2},\vec{r_2},t)
\fr{e^{2i\vec{l}\vec{r}}}{(k-l)^2 (k+l)^2}d^3r d^3l d^2k_{\bot}.
\eeq

\section{Final result for realistic beams}

The above integration was performed in the explicit form in ref. 
\cite{MSer}. The qualitative explanation of result was given by
{\em G.L.~Kotkin}. The final result is written via the
transversal size of the beam $a$ and muon time of life $\tau$ in
the form like eq. (\ref{crsec}):
\beqn 
\sigma_{eff}&=&\fr{a}{c\tau} g(x)\sigma_0; \sigma_0=20\mbox { nb};\\ 
g(x)&=&\fr{12}{5}x\sqrt{x(1-x)}\left(1+\fr{22}{9}x-\fr{16}{9}x^2\right).
\eeqn
For $a=3\mu m$, the maximal value of this "cross section"
about 0.3 fb at $\sqrt{s}\approx 100$ GeV.

Therefore, for the realistic beams, the instability in the muon
mass is invisible in the result, the effect of the finite beam
size is dominant.

\section{Conclusion}

The main conclusions are: 
\begin{itemize}
\item It is necessary to consider the distinction of the initial
state of unstable particle from the plane wave, to eliminate the
t--channel singularity discussed. Depending of the problem
considered, the dominant regularizing effect is due to either   
complex mass of unstable particle or beam sizes. There is no
theory now to consider intermediate case.
\item The discussed effect has no practical meaning for the muon 
collider.  
\end{itemize}

{\large\bf Acknowledgments}

I am very thankful to F.~Boudjema, who paid my attention to the
problem, K.~Melnikov and V.~Serbo for the information of their
result, G.~Kotkin, for the stimulating discussions.  I am
grateful to A.~Pilaftsis and T.A.~Weber for the information
about old papers in this field. Besides, I am grateful
W.~B\"uchmuller, V.~Chernyak, G.~Jikia, M.~Terentjev,
A.~Vainstein and P.~Zerwas for the discussion of the results.

\end{document}